\documentclass[preprint]{aastex631}
\usepackage{graphicx}  
\usepackage{dcolumn}   
\usepackage{bm}        
\usepackage{amssymb}   
\usepackage{upgreek}
\usepackage{mathtools}
\usepackage{hyperref}
\begin{document}

\title{Stability of the Potential Super Jupiter in Alpha Centauri System}
\author[0000-0002-3605-6597]{Tinglong~Feng}
\email{arendelle.ftl@gmail.com}
\affiliation{%
	Xi'an Jiaotong University
}%

\date{\today}

\begin{abstract}
The stability of potential exoplanets in binary star systems remains a critical challenge in celestial mechanics. This study investigates the stability of a hypothetical Jupiter-mass planet in the Alpha Centauri AB system by drawing parallels with the recently detected Neptune-mass planet in the GJ65AB system, which shares similar mass ratios and orbital eccentricities. Utilizing the Kolmogorov-Arnold-Moser (KAM) theorem and the Mean Exponential Growth of Nearby Orbits ({\tt MEGNO}) parameter within the {\tt REBOUND} software package, we perform numerical simulations to identify stable orbital configurations. Our results indicate that a Jupiter-mass planet could theoretically maintain a stable orbit in Alpha Centauri AB, with orbital parameters derived from the stable zones observed in GJ65AB. Despite the absence of observational evidence for such a planet in Alpha Centauri AB, this study suggests that similar binary systems could exhibit analogous stability characteristics. Future observational efforts are recommended to explore this potential and refine our understanding of planetary formation in binary star systems.
\end{abstract}

\section{Introduction}\label{I}
The three-body problem, which seeks stable orbit configurations among gravitating bodies, is a longstanding challenge in celestial mechanics. Recent research has prioritized specific astronomical systems, employing a range of methods including numerical simulations(\cite{10.1111/j.1365-2966.2011.19657.x}, \cite{georgakarakos2024empiricalstabilitycriteria3d}), analytical techniques (\cite{Shevchenko_2015}), and machine learning (\cite{10.1093/mnras/sty022}). 

As the closest triple stellar system to Earth, Alpha Centauri system has attracted diverse studies in astronomy, including exoplanet stability(\cite{1}, \cite{universe10030120}). Latest observations hinted at the existence of a low mass planet in Alpha Centauri (\cite{Wagner2021}), but no Jupiter-mass planet has ever been detected up to now. Given the significant distance from star C to stars A/B, the stability of planets around stars A/B can be regarded as a stability problem in a binary system(\cite{universe10030120}).

Recently, a Neptune-mass candidate planet has been detected in M-dwarf binary system GJ65(\cite{refId0}). GJ65AB and Alpha Centauri AB have almost the same mass ratio and eccentricity, from which we may suppose the planet stability for these two system may share similar property. 

In this work, we study the similarity between GJ65AB and Alpha Centauri AB, and explore the stability of a potential super Jupiter in the latter. In section \ref{II} we review the exoplanet stability theorem based on KAM theorem, from which we assume the similarity method. In section \ref{III} we compare the parameters between the two systems, and use the similarity method to derive the mass and semimajor of the potential stable planet in Alpha Centauri AB, according to parameters of the newly detected planet in GJ65AB. In section \ref{IV} we present numerical simulations to verify the stability of the newly detected super Neptune in GJ65AB and the potential super Jupiter derived from similarity method in Alpha Centauri AB. Finally, we conclude our results in section \ref{V}.

\section{Stability of Exoplanet System}\label{II}
In order to discuss the stability of exoplanet system, we first need to review some point of KAM theorem(\cite{Shikita_2010}). Consider $N$-body problem where the Hamiltonian is given by
\begin{equation}\label{1}
	H=\sum_{j=1}^{N}\frac{p_j^2}{2m_j}-\sum_{1\le j\le k\le N}G\frac{m_jm_k}{|x_j-x_k|}.
\end{equation}
It is well known that for $N\ge3$ the Hamiltonian system is not integrable. Nevertheless, it can be regarded as a small perturbation of an integrable system. If the total Hamiltonian $H(p,q)$ satisfies
\begin{equation}\label{2}
	H(p,q)=H_0(p,q)+\epsilon H_1(p,q)
\end{equation}
where $H_0$ is completely integrable, then for a sufficiently small $\epsilon$ the quasi-periodic solutions (corresponding to a stable motion in planet systems) survive under this small perturbation. That is the result of KAM theorem.

Now we turn to the stability problem for the planets in binary star system. For $N=3$ case, if we set $M_1\gg m,M_2\gg m$ and recall that Kepler problem is completely integrable, then  \eqref{2} becomes
\begin{equation}\label{3}
	H(p,q)=H_0(p,q)-G\frac{mM_1}{|x_3-x_1|}-G\frac{mM_2}{|x_3-x_2|}
\end{equation}
where $H_0$ is the Hamiltonian of binary stars $M_1,M_2$ and the kinetic energy of planet $m$ is  ignored. Compare \eqref{2} and \eqref{3} we could find that the $\epsilon$ only depends on the mass ratio of these three bodies $m:M_1:M_2$ and the relative distance of them $|x_1-x_2|:|x_2-x_3|:|x_1-x_3|$, while the latter can be described by eccentricity of orbits. Hence we could guess that for two binary star systems with nearly identical mass ratio and eccentricity, they will have similar planet stability property. For instance, consider two binary systems A$(M_1, M_2)$ and B$(M_1',M_2')$ where $M_1/M_1'=M_2/M_2'=\alpha$ and $e=e', a/a'=\beta $ (with '$a$' representing the semimajor axis). If a planet in system A with mass $m$, semimajor axis $a_3$, and eccentricity $e_3$ exhibits stable motion, then a planet in system B with mass $m/\alpha$, semimajor axis $a_3/\beta$ and eccentricity $e_3$ should also have stable motion.
\section{Methods}\label{III}
\subsection{Similarities between GJ65AB and Alpha Centauri AB}
According to the latest astrometry results(\cite{Akeson_2021}, \cite{Gaia}), GJ65AB and Alpha Centauri AB have nearly identical mass ratio and eccentricity (Table \ref{t-1}). If the former is denoted as system A and the latter as system B, as in section \ref{II}, then we have corresponding $\alpha\simeq1/9$ and $\beta\simeq1/4.2$. A novel astrometric detection result \cite{refId0} shows the existence of a stable Neptune-mass planet in GJ65 system, based on which and the hypothesis in section \ref{II} we could suppose the possibility of potential Jupiter-mass planet in Alpha Centauri AB system, at least from the standpoint of stability. The parameters of both planets are listed in Table \ref{t-1}.
\begin{table}[ht]
	\caption{The comparison of GJ65 AB and Alpha Centauri AB, parameters are from recent astrometric measurements \cite{Gaia} and \cite{Akeson_2021}. The planet's parameters in GJ65 AB are from \cite{refId0}, and the values in parentheses are derived using similarity method.\label{t-1}}
	\begin{ruledtabular}
		\begin{tabular}{lcr}
			Parameter& GJ65AB & Alpha Centauri AB \\
			\hline
			$a$ (AU) & $5.459\pm0.002$ & $23.336\pm0.013$\\
			$e$ & $0.6172\pm0.0001$ & $0.51947\pm0.00015$\\
			$\omega$ (deg) & $103.2\pm0.1$ & $231.519 \pm 0.027$\\
			$m_A(M_{\odot})$ & $0.122 \pm0.002$ & $1.0788 \pm 0.0029$\\
			$m_B(M_{\odot})$ & $0.116 \pm 0.002$ & $0.9092 \pm 0.0025$\\
			\hline
			Potential Planet Parameter: & & \\
			$a$ (AU) & $0.283 \pm 0.002$ & $(1.189)$\\
			$e$ & $0.33\pm0.30$ & $(0.33)$\\
			$m(M_{\oplus})$& $40$& $(350)$\\
		\end{tabular}
	\end{ruledtabular}
\end{table}
\subsection{Numerical Simulation}
The similarities between GJ65AB and Alpha Centauri AB, together with the newly detected stable super Neptune in GJ65 system, suggest the stability of the  corresponding potential super Jupiter in Alpha Centauri AB. In order to verify this, we employ the Mean Exponential Growth of Nearby Orbits ({\tt MEGNO}) stability parameter(\cite{MEGNO}). 

The {\tt MEGNO} parameter evaluate the stability of N-body gravitational system by measuring how orbits in phase space  diverge over time. In {\tt MEGNO}, values near two indicate stable orbits, while significantly higher values suggest chaos and instability. For this study, we used the {\tt MEGNO} tool from the {\tt REBOUND} software package(\cite{REBOUND}).

Here is our experimental approach: first, in GJ65 system, we placed the planet with mass and orbit parameters following the newly astrometric detection result so that we can verify the stability of this exoplanet. Then, in Alpha Centauri AB we placed the potential planet with corresponding parameters in Table \ref{t-1}, derived from similarity method, to verify whether it is in stable region. 
\section{Results}\label{IV}
\begin{figure}
	\centering
	\includegraphics[width=\textwidth]{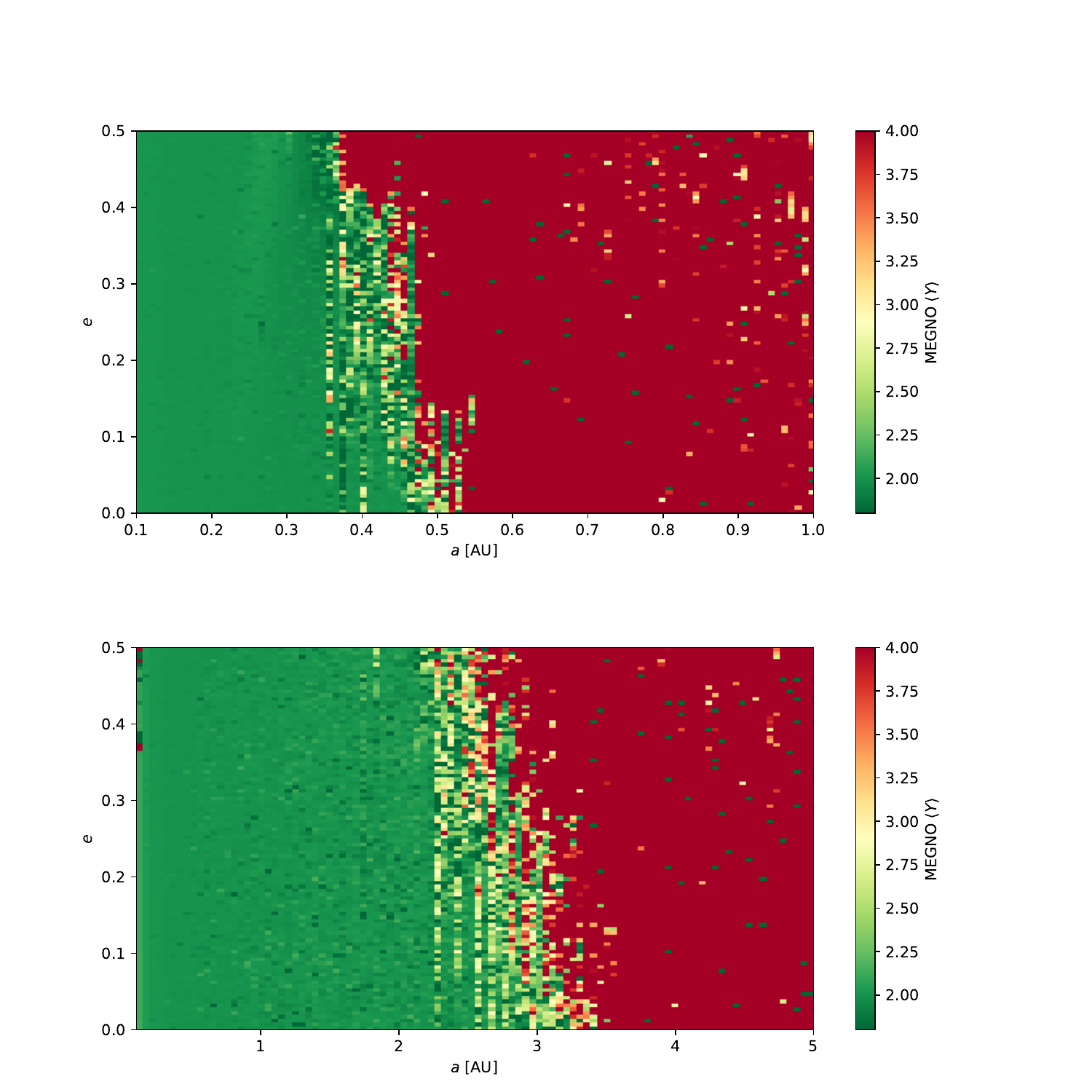}
	\caption{Stability maps based on {\tt MEGNO} values for Neptune-mass planet in GJ65AB(\textbf{top panel}) and for Jupiter-mass planet in Alpha Centauri AB(\textbf{below panel}). Dynamically stable regions are coloured in green.\label{fig-1}}
\end{figure}

First we perform the simulation of planet in GJ65 system. Using the most recent orbital solution based on astrometric measurements, we initialized the binary orbital parameters with the following values: semimajor axis $a=5.459$ AU, eccentricity $e=0.6172$, and argument of periastron $\omega=103.2$ deg. The masses were set to $m_A=0.122 M_{\odot}$ and $m_B=0.116 M_{\odot}$, respectively. For simplicity, we supposed the host star of the planet is star A, and set the mass of the planet $40 M_{\oplus}$, according to recent detection. For this simulation, we restricted the semimajor axis of the planet to range from $0.1$ to $1.0$ au, and eccentricities less than $0.5$.

The top panel in Figure \ref{fig-1} shows the result of the simulation. The green regions indicate the stable zones in the $a-e$ parameter space, while the red regions are chaotic zones. We identified the stable zone spanning from $0.1$ to $\sim0.35$ au, which contains all the stable orbits for $e $ ranging from $0$ to $0.5$. The astrometric detection suggests the planet has the following parameters: $a\approx0.283$ au and $e\approx0.33$, which are indeed located in the stable zones of our simulation, and the stability of this planet has been verified.

Next we turn to Alpha Centauri AB. According to highly precise astrometric results, we set the parameters: $a=23.336$ AU, $e=0.5195$, $\omega=231.5$ deg, $m_A=1.0788 M_{\odot}$ and $m_B= 0.9092M_{\odot}$. The potential planet was also supposed to be around star A, and the planet's mass was set to be about $350M_{\oplus}$ based on similarity method, i. e., a Jupiter mass exoplanet. For this simulation, we restricted the semimajor axis of the planet to range from $0.1$ to $5.0$ au, and eccentricities less than $0.5$.

The panel below in Figure \ref{fig-1} displays the result of this simulation, where green zones indicate stability and red ones represent chaos. We figured out the stable zone, with $a$ spanning from $0.1$ to $\sim2.2$ au and $e$ ranges from $0$ to $0.5$. According to the similarity discussed at the beginning of this section, the potential stable planet should have parameters of $a\approx1.189$ and $e\approx0.33$, exactly contained in the stable zone of our simulation. This result verified our postulate that, if there exists a Jupiter-mass planet in Alpha Centauri AB system with corresponding orbital parameters, it should be stable.

\section{Conclusion}\label{V}
In this work, we propose that exoplanets in similar binary systems, i.e., systems with nearly identical mass ratios and eccentricities, may exhibit similar stability properties according to the KAM theorem. From this hypothesis, together with the newly detected Neptune-mass planet in GJ65 system which is similar to Alpha Centauri AB, we assume the existence of a potential Jupiter-mass planet with corresponding orbital parameters in Alpha Centauri AB should also be possible, at least from the standpoint of dynamical stability.  We have used the numerical tool {\tt MEGNO} to successfully verify our assumption. 

Nevertheless, we have not detected any Jupiter-mass planet candidates in Alpha Centauri AB system up to now(\cite{Wagner2021}). To explain this we may come up with two possible cases:
\begin{enumerate}
	\item There exists Jupiter-mass planet in Alpha Centauri AB, but we have not detected it yet. 
	\item There does not exist a Jupiter-mass planet in  Alpha Centauri AB. This case suggests the formation of giant planets in binary systems depends not only on the mass ratio and eccentricity but also on the absolute value of the host star’s mass.
\end{enumerate}
These two cases may indicate further studies and detection.
\bibliographystyle{aasjournal}
\bibliography{ref}
\end{document}